\documentclass[11pt]{article}
\usepackage{amssymb,amsmath,cite}
\catcode`\@=11

\@addtoreset{equation}{section}
\def\theequation{\arabic{section}.\arabic{equation}}
     
\catcode`\@=11
\def\thesection{\arabic{section}}

\def\appendix{\setcounter{section}{0}
        \def\thesection{Appendix.}
        \def\theequation{\Alph{section}.\arabic{equation}}}
\def\section{\@startsection{section}{1}{\z@}{3.5ex plus 1ex minus
   .2ex}{2.3ex plus .2ex}{\large\bf}}

\long\def\@makefntext#1{\parindent 0cm\noindent
\hbox to .5em{\hss$^{\@thefnmark}$}#1}

\newcommand{\G}{{\cal G}}
\newcommand{\HH}{{\cal H}}

\begin{document}
\begin{titlepage}
\vspace{.5in}
\begin{flushright}
ESI-2150\\  
UCD-09-04\\
June 2009\\  
\end{flushright}
\vspace{.5in}
\begin{center}
{\Large\bf
 Chiral Topologically Massive Gravity\\[1ex]
and Extremal B-F Scalars}\\  
\vspace{.4in}
{S.~C{\sc arlip}\footnote{\it email: carlip@physics.ucdavis.edu}\\
       {\small\it Department of Physics}\\
       {\small\it University of California}\\
       {\small\it Davis, CA 95616}\\{\small\it USA}}
\end{center}

\vspace{.5in}
\begin{center}
{\large\bf Abstract}
\end{center}
\begin{center}
\begin{minipage}{4.8in}
{\small
At a critical ``chiral'' coupling, topologically massive gravity with a 
negative cosmological constant exhibits several unusual features, 
including the emergence of a new logarithmic branch of solutions and 
a linearization instability for certain boundary conditions.  I show 
that at this coupling, the linearized theory may be parametrized by 
a free scalar field at the Breitenlohner-Freedman bound, and use this 
description to investigate these features.\\[3ex]
}
\end{minipage}
\end{center}
\end{titlepage}
\addtocounter{footnote}{-1}

\section{Some puzzles}

Topologically massive gravity \cite{DJT}---three-dimensional Einstein 
gravity supplemented by a gravitational Chern-Simons term in the action%
---provides an interesting playground in which to explore quantum gravity.
In contrast to pure Einstein gravity in three dimensions, topologically
massive gravity contains a propagating physical degree of freedom.  It 
nevertheless appears to be renormalizable \cite{DeserYang,Keszthelyi,Oda}, 
perhaps offering a rare instance of a theory of spacetime geometry that 
can be treated by methods of conventional quantum field theory.

With the addition of a negative cosmological constant, topologically 
massive gravity also provides a new realm in which to investigate the
AdS/CFT correspondence \cite{Kraus,Solodukhin,LSS}.  In this
setting, though, the theory has a dangerous instability: the local degree 
of freedom contributes to the energy with a sign opposite to that of the 
BTZ black hole \cite{Moussa}, suggesting the absence of a ground state.  
Using the supersymmetric extension of topologically massive gravity, 
one can prove a positive energy theorem \cite{Deser,Abbott}, but only 
in the sector containing no black holes.   

As Li, Song, and Strominger have noted, however, the structure of the
theory changes dramatically at a special ``chiral'' value of the coupling
constants \cite{LSS}.  At this coupling, the propagating ``massive 
graviton'' modes become pure gauge, and can be eliminated by
diffeomorphisms \cite{Strominger}.  This presents a puzzle, since
the constraint analysis shows the continued existence of a local degree 
of freedom \cite{Carlip,HMT}.  The answer is partially understood 
\cite{GJ,GKP}: precisely at the chiral coupling, a new set of local 
``logarithmic'' modes appears in the linearized theory.\footnote{The
appearance of logarithmic modes in \emph{exact} pp-wave solutions 
at the chiral coupling was noted earlier in \cite{Ayon}.}  These modes  
violate the Brown-Henneaux boundary conditions \cite{BH} usually 
imposed in (2+1)-dimensional asymptotically AdS gravity, and it has 
been argued that they can be eliminated by imposing suitable boundary 
conditions.  If so, topologically massive gravity at the chiral coupling 
would become a truly chiral theory, closely related to a chiral half of 
ordinary Einstein gravity. 

But this leads to another puzzle.  As demonstrated in \cite{CDWW}, the 
logarithmic modes at the chiral coupling can be generated, at least in 
the linearized theory, from strictly local initial data.  The imposition 
of Brown-Henneaux boundary conditions then becomes a teleological 
choice, a restriction on initial data ``now'' on the basis of its behavior 
in the future.  A possible resolution, proposed in \cite{MSS}, could come 
from a linearization instability: although linearized initial data may
have compact support, there is evidence that at the next order in 
perturbation theory, such data may violate suitably strong boundary 
conditions.

Any resolution of these puzzles requires an understanding of the 
interplay between boundary conditions and nonlinearity in topologically 
massive gravity.   In this paper, I find the  general solution to 
the linearized field equations in a Poincar{\'e} coordinate patch, and 
show that it can be parametrized by solutions---or initial data---of a 
free scalar field whose mass lies at the Breitenlohner-Freedman bound 
\cite{BF}.  I then use this parametrization to investigate the questions 
of boundary behavior and linearization stability.

\section{Linearized topologically massive gravity and B-F scalars
\label{s1}}

Consider a free scalar field $\varphi$ with mass $m$ in three-dimensional 
anti-de Sitter space.  In a Poincar{\'e} coordinate patch\footnote{A 
Poincar{\'e} patch does not cover the whole of AdS, but we will be 
interested in local initial data, for which such a coordinate choice is 
sufficient and greatly simplifies computations.} (with units $\ell=1$), 
the metric is
\begin{align}
d{\bar s}^2 = \frac{1}{z^2}\left( 2dx^+dx^- + dz^2\right) 
\label{a1}
\end{align}
with $x^\pm = \frac{1}{\sqrt{2}}(x\pm t)$, and the Klein-Gordon equation 
is simply
\begin{align}
\left[ 2z^2\partial_+\partial_- + (N-1)^2 - (m^2+1)\right]\varphi = 0 ,
\label{a2}
\end{align}
where $N=z\partial_z$.  Near the conformal boundary $z=0$, solutions 
behave as
\begin{align}
\varphi \sim z^{1\pm\sqrt{1+m^2}} .
\label{a3}
\end{align}
One solution thus seems to disappear at the Breitenlohner-Freedman bound 
$m^2=-1$.  This behavior is well-understood, though: 
precisely at $m^2=-1$, a new solution appears with an asymptotic behavior
\begin{align}
\varphi \sim z\ln z .
\label{a4}
\end{align}

This is exactly the behavior of linearized topologically massive 
gravity noted by Grumiller and Johansson  at the chiral coupling 
\cite{GJ}.  Indeed, the two theories are intimately related.   Let 
$\varphi$ be an extremal B-F scalar, that is, one satisfying (\ref{a2}) 
with $m^2=-1$, and define $X = \varphi/z^3$, so
\begin{align}
\left[ 2z^2\partial_+\partial_- + (N+2)^2\right]X= 0 .
\label{a5}
\end{align}
Consider a linear excitation $g_{\mu\nu} = {\bar g}_{\mu\nu} 
+ h_{\mu\nu}$ of the metric around AdS space, and choose Gaussian 
normal coordinates in $z$,
\begin{align}
h_{+z} = h_{-z} = h_{zz} = 0 ,
\label{a6}
\end{align}
as in the Fefferman-Graham expansion \cite{FG}.   As I show in the 
appendix, the perturbation
\begin{align}
h_{++} &= -2\partial_+^2(N+2)X \nonumber \\
h_{+-} &= 2\partial_+\partial_-(N+2)X = -\frac{1}{z^2}N(N+2)^2X 
  \label{a7} \\
h_{--} &= -2\partial_-^2(N-2)X \nonumber
\end{align}
then exactly solves the linearized field equations of 
topologically massive gravity at the chiral coupling, thus parametrizing 
solutions by a scalar at the B-F bound.

More than that: as demonstrated in the appendix, the \emph{general} 
solution in a Poincar{\'e} coordinate patch can be obtained by adding 
to (\ref{a7}) a solution of the form
\begin{align}
{\tilde h}_{++} &=  \frac{\partial_+{}^2c}{z^2} 
                  + \frac{1}{4}\partial_+{}^3\partial_-(2a-b-c)  
                  + \gamma^+\ln z \nonumber\\
{\tilde h}_{+-} &=  \frac{\partial_+\partial_-a}{z^2} 
                  + \frac{1}{4}\partial_+{}^2\partial_-{}^2(2a-b-c)
\label{a8}\\
{\tilde h}_{--} &=  \frac{\partial_-{}^2b}{z^2} 
                  + \frac{1}{4}\partial_+\partial_-{}^3(2a-b-c) , \nonumber
\end{align}
where $a$, $b$, and $c$ are arbitrary functions of $x^+$ and $x^-$ and 
$\gamma^+$ depends only on $x^+$.  The linearized curvature of $\tilde h$ 
depends only on $\gamma^+$, so the remaining piece of $\tilde h$ is 
formally equivalent to a diffeomorphism; in fact, it is the most general 
diffeomorphism that preserves the gauge condition (\ref{a6}) at first 
order.

The general solution of linearized topologically massive gravity at the
chiral coupling in a Poincar{\'e} coordinate patch thus consists of 
three pieces: a dynamical degree of freedom parametrized by an extremal 
B-F scalar; a formal diffeomorphism (that need not satisfy the boundary 
conditions to make it ``pure gauge''); and a chiral logarithmic term 
$h_{++}=\gamma^+(x^+)\ln z$ corresponding to the pp-wave of
\cite{CDWW,Ayon}.

\section{Boundary values \label{s2}}

At an initial time, the scalar field $X$ may be chosen to have compact 
support away from the conformal  boundary.  Indeed, we can specify 
arbitrary initial data for $X$ and $\partial_t X$ and evolve it forward 
with the Klein-Gordon equation.  As the system evolves, though, the field 
will reach the conformal boundary in a finite time, and we may ask about 
compatibility with boundary conditions.  Note that we risk making a 
teleological argument here, that is, limiting initial data ``now'' on 
the basis of what the field will do in the future.   

Anti-de Sitter space is not globally hyperbolic, so initial data  
is not enough to determine a unique evolution.  For the case of a free 
scalar, however, Ishibashi and Wald have shown that there is a 
finite-dimensional family of ``nice'' evolution laws, where niceness 
includes the existence of a positive energy, compatibility with time 
translation symmetry, and a set of convergence conditions \cite{IshiWald}. 
In particular, for a scalar field $\varphi=z^3X$ at the Breitenlohner-%
Freedman bound, the asymptotic behavior near
$z=0$ is
\begin{align}
\varphi \sim a_0z + b_0 z\ln z + \dots ,
\label{b1}
\end{align}
and the allowed evolution laws correspond uniquely to choices of the
ratio $a_0/b_0$.  (Positivity imposes an interesting limit on this ratio, 
whose implications have not, as far as I know, been explored in the 
context of topologically massive gravity.)

Each choice of an evolution law gives rise to a Greens function, and in 
principle one could use such a function to determine the future boundary 
behavior of  $X$.  Let us take a shortcut, and directly evaluate $X$---and 
hence $h_{\mu\nu}$---near $z=0$, keeping in mind that we are really 
looking at the future evolution of our compact initial data.  Since 
$X=\varphi/z^3$, the appropriate form of (\ref{b1}) is now
\begin{align}
X \sim \frac{\alpha_0}{z^2} + \frac{\beta_0\ln z}{z^2} 
   + \alpha_1 + \beta_1\ln z +\dots ,
\label{b2}
\end{align}
where the coefficients are functions of $x^+$ and $x^-$.  The equations 
of motion (\ref{a5}) then yield
\begin{align}
\alpha_1 = \frac{1}{2}\partial_+\partial_-(\beta_0-\alpha_0), \qquad
\beta_1 = -\frac{1}{2}\partial_+\partial_-\beta_0 .
\label{b3}
\end{align}
This, in turn, implies from (\ref{a7}) that
\begin{align}
h_{++} &\sim -\frac{2\partial_+{}^2\beta_0}{z^2} 
     -\partial_+{}^3\partial_-(\beta_0-2\alpha_0) 
     + 2\partial_+{}^3\partial_-\beta_0\ln z  + \dots \nonumber\\
h_{+-}&\sim \frac{2\partial_+\partial_-\beta_0}{z^2} 
     +\partial_+{}^2\partial_-{}^2(\beta_0-2\alpha_0) 
      - 2\partial_+{}^2\partial_-{}^2\beta_0\ln z  + \dots 
\label{b4} \\
h_{--}&\sim -\frac{2\partial_-{}^2(\beta_0-4\alpha_0)}{z^2} 
     + \frac{8\partial_-{}^2\beta_0\ln z}{z^2}
     +\partial_+\partial_-{}^3(3\beta_0-2\alpha_0) 
      - 2\partial_+\partial_-{}^3\beta_0\ln z + \dots \nonumber
\end{align}

These asymptotics clearly violate standard boundary conditions, including 
both the original Brown-Henneaux boundary conditions for (2+1)-dimensional 
gravity \cite{BH} and the weaker logarithmic boundary conditions of 
\cite{HMT,GJb}.  We have not yet looked at the general solution, however; that 
is, we have not yet added in a term of the form (\ref{a8}).  By inspection,
we can cancel the $1/z^2$ terms in (\ref{b4}) by choosing $c = 2\beta_0$, 
$a = -2\beta_0$, and $b  = 2\beta_0 - 8\alpha_0$.  The $\ln z/z^2$ term 
in $h_{--}$ cannot be canceled, though, and must be eliminated by imposing 
the condition
\begin{align}
\partial_-{}^2\beta_0 = 0 .
\label{b6}
\end{align}
It is then easy to check that 
\begin{align}
h_{++} &\sim A(x^+,x^-)  + B(x^+)\ln z +\dots \nonumber\\
h_{+-} &\sim 0 + \dots \label{b7}\\
h_{--} &\sim 0 + \dots  \nonumber
\end{align}
consistent with the logarithmic asymptotics of \cite{HMT,GJb}.  

At first sight,  the condition (\ref{b6}) appears to be a teleological 
restriction on our initial data.   As noted earlier, however, a scalar 
field in anti-de Sitter space does not have a unique Greens function.  
In particular, one can choose a Greens function for which the coefficient
$b_0$ in (\ref{b1}) vanishes for all initial data.  If we assume   
``nice'' evolution of $X$  is in one-to-one correspondence with 
``nice'' evolution of $h_{ij}$---I will return to this assumption in the 
conclusion---then the corresponding choice eliminates $\beta_0$ in
(\ref{b4}).\footnote{This choice does not eliminate the $\ln z$ 
asymptotics in (\ref{b7}), since the relevant term reappears as 
$\gamma^+$ in (\ref{a8}).}

With the proper choice of a Greens function, compact initial data
for $X$ thus evolves in a way that respects natural Brown-Henneaux
or logarithmic boundary conditions.  This is still a  delicate issue, 
though, since the cancellation of the $1/z^2$ terms in (\ref{b4})
required a ``diffeomorphism mode'' (\ref{a8}) which did not itself 
have compact support.  A direct Greens function computation of the 
evolution of the full metric perturbation would thus still be valuable.

\section{Linearization stability}

So far, we have only considered the linear approximation to 
topologically massive gravity.  For the most part, physicists are used 
to situations in which a first-order solution of a set of equations can 
be extended to at least a perturbative solution of the full nonlinear 
equations.  Sometimes, however, such an extension fails: a solution 
of a linearized set of field equations may not be the linearization of 
an exact solution.  In such a case, the theory is said to have a 
linearization instability \cite{Fischer}.

As a simple example, consider the equation $x^2=0$.  If we expand 
around ${\bar x}$, the linearized equation is ${\bar x}\delta x = 0$; 
thus if we choose ${\bar x}=0$, $\delta x$ is unconstrained at linear 
order, although only $\delta x = 0$ is a linearization of the exact 
solution $x=0$.  While the instances in physics are more complicated, 
this example illustrates the fundamental point: if one expands around 
too special a background, some of the linearized equations may vanish 
identically, and the first constraints may be quadratic.

In ordinary general relativity, a linearization instability can occur 
if one expands around a background metric admitting Killing vectors.  
Recall that in $d$ dimensions, $d$ of the Einstein field equations are 
constraints; that is, when smeared against a vector $\xi^\mu$, they 
generate diffeomorphisms $g_{\mu\nu}\rightarrow g_{\mu\nu} 
+ \nabla_\mu\xi_\nu + \nabla_\nu\xi_\mu$.  But if $\xi_\mu$ is a 
Killing vector for a background metric ${\bar g}_{\mu\nu}$, this 
transformation is  trivial at first order, and up to possible boundary 
terms, the corresponding generators vanish.  In other words, if the 
background metric admits a Killing vector, certain combinations of the 
field equations first appear at second order.  With particular boundary 
conditions, these second-order field equations may then restrict the 
allowed first-order perturbations; see \cite{Brill,Higuchi} for 
simple examples. 

Maloney, Song, and Strominger have argued in \cite{MSS} that such a 
phenomenon occurs for topologically massive gravity at the chiral 
coupling.  Their analysis used the nonlocal light front modes of 
\cite{CDWW}, making the computations rather complex.  The present
B-F scalar formalism allows a straightforward check of their results.

We begin with some simple manipulations of the exact field equations.  
Define
\begin{align}
\G_{\mu\nu} = R_{\mu\nu} - \frac{1}{2}g_{\mu\nu}R - g_{\mu\nu} .
\label{c1}
\end{align}
The full field equations of topologically massive gravity at the chiral coupling 
are then
\begin{align}
E_{\mu\nu} = \G_{\mu\nu} 
  + \frac{1}{2}\left( \epsilon_\mu{}^{\alpha\beta}\nabla_\alpha\G_{\beta\nu}
  + \epsilon_\nu{}^{\alpha\beta}\nabla_\alpha\G_{\beta\mu}\right)  = 0  .
\label{c2}
\end{align}
Let $\xi^\mu$ be an arbitrary vector and define
\begin{align}
\Delta_{\alpha\beta}[\xi] 
   = \nabla_\alpha\xi_\beta + \nabla_\beta\xi_\alpha, \qquad
\chi^\rho[\xi] 
   = -\xi^\rho - \frac{1}{2}\epsilon^{\rho\alpha\beta}\nabla_\alpha\xi_\beta .
\label{c3}
\end{align}
$\Delta_{\alpha\beta}[\xi]$ measures the failure of $\xi^\mu$ to be a 
Killing vector; $\chi^\mu[\xi]$ measures the chirality of $\xi^\mu$, in the 
sense that in anti-de Sitter space, 
$$
\chi^\mu = \left\{\begin{array}{cl} 
                 0 & \hbox{if\ $\xi^\mu = \delta^\mu_+$}\\[.5ex]
                 -\delta^\mu_z & \hbox{if\ $\xi^\mu = \delta^\mu_z$}\\[.5ex]
                 -2\delta^\mu_- & \hbox{if\ $\xi^\mu = \delta^\mu_-$}
\end{array}  \right.
$$
As I show in the appendix, it is then an exact equality that
\begin{align}
\xi^\mu E_\mu{}^\nu =  -\chi^\mu\G_\mu{}^\nu 
    + \frac{1}{2}\epsilon^{\nu\alpha\beta}\Delta_{\mu\beta}\G_\alpha{}^\mu
    +\nabla_\alpha{\cal F}^{\alpha\nu}  
\quad\hbox{with}\ 
{\cal F}^{\alpha\nu}  
   = \epsilon^{\nu\alpha\beta}\xi_\mu\left(\G_\beta{}^\mu 
    - \frac{1}{2}\delta^\mu_\beta\G\right) .
\label{c4}
\end{align}
This is of the right form for a linearization instability: if one expands 
around anti-de Sitter space, for which ${\bar\G}_\mu{}^\nu=0$, the 
first-order ``bulk'' contribution to the smeared field equation
\begin{align}
I_\xi = \int d^2x\sqrt{|g|}\,\xi^\mu E_\mu{}^t = 0
\label{c6}
\end{align}
vanishes if $\xi^\mu$ is an AdS Killing vector (${\bar\Delta}_{\mu\nu}=0$) 
of the proper chirality (${\bar\chi}^\mu=0$).  Suppose in addition that 
boundary conditions force the  contribution from $\cal F^{\alpha\nu}$ 
to vanish.  Then, as above, a particular combination of the equations of 
motion is identically zero at first order, and new restrictions may be 
expected at second order.  

To see whether such restrictions really occur, we must evaluate the 
second-order contribution $I_\xi^{(2)}$.  Let us choose $\xi^\mu =
\delta^\mu_+$.   Then $\Delta_{\mu\nu}$, $\chi^\mu$, and $\G_\mu{}^\nu$ 
all vanish at zeroth order, so from (\ref{c4})
\begin{align}
\sqrt{|g|}\,\xi^\mu E_\mu^{(2)+} 
   &= -z^2(\sqrt{|g|}\,\chi^\mu)^{(1)}\HH_{\mu -} 
   + \frac{z^2}{2}\left(\Delta_{+-}^{(1)}\HH_{z-} 
   + \Delta_{--}^{(1)}\HH_{z+}\right)
   + \partial_\alpha(z^{-3}{\cal F}^{(2)\alpha+}) \nonumber\\
\sqrt{|g|}\,\xi^\mu E_\mu^{(2)-}  
   &= -z^2(\sqrt{|g|}\,\chi^\mu)^{(1)}\HH_{\mu +} 
   - \frac{z^2}{2}\left(\Delta_{++}^{(1)}\HH_{z-} 
   + \Delta_{+-}^{(1)}\HH_{z+}\right)
   + \partial_\alpha(z^{-3}{\cal F}^{(2)\alpha-})  ,
\label{c7}
\end{align}
where $\HH_{\mu\nu} = \G_{\mu\nu}^{(1)}$ is the first-order curvature.  
Moreover, from (\ref{c3}),
\begin{align}
(\sqrt{|g|}\,\chi^{\mu})^{(1)} = -\frac{1}{z}h_{+-}\xi^\mu
   - \frac{1}{2}{\tilde\epsilon}^{\mu\alpha\beta}\partial_\alpha h_{\beta+}, 
   \quad
\Delta_{\mu\nu}^{(1)} = \partial_+h_{\mu\nu}.  
\label{c7a}
\end{align}

Now, recall from section \ref{s2} that a general solution $h_{ij}$ 
of the linearized field equations consists of three pieces, one determined 
by an extremal B-F scalar, one formally equivalent to a diffeomorphism, 
and one a chiral pp-wave depending on $\ln z$.  Of these, only the first 
permits arbitrary initial data of compact support, so let us begin with 
that piece.  I show in the appendix that 
\begin{align}
\sqrt{|g|}\,\xi^\mu E_\mu^{(2)t}[X] = 
   &-\frac{1}{\sqrt{2}}\frac{1}{z}\left[ \left(\partial_zN(N+2)^2X\right)^2
   + 2\left(\partial_+ N(N+2)^2X\right)^2\right] \nonumber\\
   &+ \partial_\alpha(z^{-3}{\cal F}^{(2)\alpha t}) 
   + \hbox{\it spatial boundary terms} ,
\label{c8}
\end{align}
where the boundary terms vanish if $X$ has compact support. The integral 
$I_\xi$ of (\ref{c6}) is thus
\begin{align}
I_\xi^{(2)} = -\frac{1}{\sqrt{2}}\int dx\,dz\,\frac{1}{z}\left[ 
   \left(\partial_zN(N+2)^2X\right)^2
   + 2\left(\partial_+N(N+2)^2X\right)^2\right] 
   + \int_{z=0}\!\!dx\, z^{-3}{\cal F}^{(2)zt} .
\label{c9}
\end{align}
In general, one must also consider terms involving the remaining two 
pieces in ${\tilde h}_{ij}$, but I show in the appendix that these give 
only a boundary term that vanishes if $X$ has compact support.

In accord with the results of \cite{MSS}, we see that the ``bulk'' term 
in (\ref{c9}) is negative definite.  The ``boundary'' term is precisely 
the Abbott-Deser-Tekin charge $Q^+$ \cite{Abbott,DT}.  For the general 
logarithmic boundary conditions of \cite{HMT,GJb},
\begin{align}
h_{++} &\sim  A_{++}(x^+,x^-) + B(x^+)\ln z + \dots \nonumber\\
h_{+-} &\sim   A_{+-}(x^+,x^-) + \dots  \label{c11}\\
h_{--} &\sim A_{--}(x^-)  + \dots, \nonumber
\end{align}
it is easily checked that 
\begin{align}
Q^+ = -\frac{1}{\sqrt{2}}\int_{z=0}\!\!dx\, B .
\label{c12}
\end{align}
By the equations of motion, the total integral $I_\xi$ must vanish.  Thus 
if our perturbation expansion remains valid throughout the Poincar{\'e} 
patch (so that we can separately require that $I_\xi^{(2)}=0$) and if we 
can consistently restrict ourselves to Brown-Henneaux boundary conditions 
$B(x^+) = 0$, the bulk integrand in (\ref{c9}) must be zero:
\begin{align}
\partial_zN(N+2)^2X = \partial_+N(N+2)^2X = 0 .
\label{c13}
\end{align}
The general solution of (\ref{a5}) and (\ref{c13}) is
\begin{align}
X &= \frac{\alpha_0}{z^2} + \frac{\beta_0\ln z}{z^2} 
   + \frac{1}{2}\partial_+\partial_-(\beta_0-\alpha_0) 
   +  \frac{1}{2}\partial_+\partial_-\beta_0\ln z  
   \quad \text{with $\partial_+{}^2\partial_-{}^2\alpha_0 
   = \partial_+{}^2\partial_-\beta_0 = 0$} .
\label{c14}
\end{align}
The resulting linearized curvature has nonvanishing components
\begin{align} 
\HH_{--} = \frac{8\partial_-{}^2\beta_0}{z^2} 
                  + 2\partial_+\partial_-^3\beta_0 , \qquad
\HH_{-z} = \frac{4\partial_+\partial_-{}^2\beta_0}{z} .
\label{c16}
\end{align}
In particular, if we impose any reasonable AdS boundary conditions, we 
must require that $\partial_-{}^2\beta_0=0$, which in turn implies 
that the entire linearized curvature tensor is zero.  

We thus confirm the results of \cite{MSS}: if our perturbation 
expansion remains valid at $z=0$, and if the sector $Q^+=0$ can 
be consistently treated as a superselection sector, then this sector 
exhibits a linearization instability, and only those linearized solutions 
with vanishing curvature can be extended to second order.  Equivalently, 
if a metric perturbation with nonvanishing $\HH_{\mu\nu}$ has 
compact support at first order, the field equations force the second 
order perturbation to have a nonvanishing boundary contribution to 
the charge $Q^+$, and hence a nonvanishing logarithmic term in
the asymptotic expansion (\ref{c11}).  If such a logarithmic term can
be consistently forbidden by boundary conditions, such first order
metric perturbations are thus excluded.

Finally, let us note one more connection between chiral topologically 
massive gravity and scalar fields.  The integral (\ref{c9}) that 
controls linearization instability depends on $X$ only through
\begin{align}
\phi = \sqrt{2} N(N+2)^2X .
\label{c17}
\end{align}
It is easily checked from (\ref{a5}) that $\phi$ obeys the equations of 
motion for a \emph{massless} scalar field.  Furthermore, the ``bulk'' 
term in (\ref{c9}) is simply the stress-energy tensor for $\phi$:
\begin{align}
-\frac{1}{\sqrt{2}} \frac{1}{z}\left[ \left(\partial_zN(N+2)^2X\right)^2
   + 2\left(\partial_+N(N+2)^2X\right)^2\right] 
= \sqrt{|g|}T_+{}^t[\phi] .
\label{c18}
\end{align}
It was noted in \cite{CDWW} that topologically massive gravity in light 
front coordinates can be described in terms of a single massless scalar 
field, at the expense of allowing nonlocal dependence of $h_{\mu\nu}$ 
on the field.  We now see that the same is true in the present setting.

\section{Where we stand}

The heart of this paper has been a demonstration that the linearized
solutions of topologically massive AdS gravity at the chiral coupling
may be parametrized by a free scalar field at the Breitenlohner-%
Freedman mass bound.  The peculiarities of the chiral coupling---in
particular, the appearance of logarithmic solutions---reflect the 
behavior of such an extremal scalar.  While light front gauge provides
an alternative scalar parametrization \cite{CDWW}, the present form 
has the advantage of locality: the metric perturbations are now strictly 
local functions of the scalar $X$.  Although this work has been carried 
out in a single Poincar{\'e} coordinate patch, there are no obvious 
obstructions to a similar construction in global coordinates.  A scalar 
field at the B-F bound also appears in the description of exact pp-waves 
at the chiral coupling \cite{Ayonb}; a further exploration of this 
relationship could be interesting.

The importance of this new parametrization, of course, depends on 
what it can tell us about the physical puzzles of chiral topologically 
massive gravity.  As we have seen, we can write down the general solution
of the linearized equations of motion, and use this to confirm the claim
of \cite{MSS} that the $Q^+=0$ sector has a linearization instability
that excludes the propagating bulk modes.  At the same time, the 
positive energy theorem of \cite{Deser,Abbott} suggests that another 
superselection sector may exist, in which black holes are excluded.  We 
are thus left with a picture of a theory containing three sectors: a 
chiral ($Q^+=0$) sector with black holes but no bulk modes, requiring 
$G>0$ for positive energy; a sector with bulk modes but no black holes, 
requiring $G<0$ for positive energy; and a larger sector with the  
logarithmic boundary conditions of \cite{HMT,GJb}, in which the energy 
may not be bounded below for any $G$.

To show this picture is correct, though, more work is needed.  First,
we do not know that the potential positive energy sectors are genuine
superselection sectors that really decouple from the rest of the 
theory.   In the chiral sector, the remaining linearized excitations  
are formal diffeomorphisms, which extend to exact solutions that are 
nontrivial only at the conformal boundary.  Until the boundary dynamics 
is better understood, however, we cannot exclude the possibility that 
consistent interactions require that the $Q^+\ne0$ ``bulk'' modes be 
present at the boundary as well.  In the ``no black holes'' sector, it 
is plausible that the positive energy bulk excitations cannot collapse 
to form negative energy black holes,  but I know no proof that
such processes are excluded.  In the ``logarithmic'' sector, we know
that energies are unbounded below at low orders of perturbation theory,
but not whether nonperturbative bounds exist.

Indeed, apart from investigations of the constraints and of solutions
with special symmetries, work on this model has relied almost  
exclusively on perturbation theory.  Efforts to 
prove, or disprove, a global positive energy theorem have not yet 
succeeded at the chiral coupling \cite{Szegin}, and we cannot exclude 
the possibility of nonperturbative surprises.  In particular, the low 
order analysis presented here, including the analysis of linearization 
instability, depends on the assumption that our perturbative 
expansion remains valid all the way out to $z=0$, despite the 
presence of terms in the expansion with inverse powers of $z$.  In 
some crude sense, we know this is not correct: if we choose a 
first-order solution with compact initial data, the second-order
solution is generically nonzero all the way out to $z=0$, so near the
boundary $h^{(2)}$ dominates $h^{(1)}$.  At this order, this 
phenomenon is merely another indication of linearization instability,
and it is plausible that it does not extend to higher orders, but a more
careful and rigorous analysis is clearly needed.

Finally, the failure of asymptotically anti-de Sitter space to be globally
hyperbolic has some subtle implications that may not be fully appreciated.
As in a globally hyperbolic spacetime, one may start with initial data 
in a compact region and evolve it forward in time with a Greens function 
of one's choice to obtain a solution of the field equations near the
initial time slice.  In a non-globally hyperbolic spacetime, however,
the resulting solution may not depend continuously on the initial data; 
that is, arbitrarily small changes in initial data may lead to large 
changes in the solution, invalidating any perturbative expansion
\cite{Schleich}.\footnote{See Ref.\ \cite{Courant},  chapter III \S6 
for a simple example of how this can occur for an elliptic differential 
equation.}  For a scalar field in anti-de Sitter space, Ishibashi and 
Wald have shown that this problem can be circumvented with a suitable 
choice of Greens functions \cite{IshiWald}, and I have used this result 
to determine the evolution of the scalar field $X$.  But while the metric 
perturbations depend locally on $X$, the converse is not true, and it 
is not obvious that the evolution described here will always depend
continuously on the initial data.  For the full nonlinear theory, the 
situation is even less clear.  Again, a much more careful and rigorous 
analysis is needed before we can be truly confident of any conclusions.

\appendix
\section{Details of some calculations}

In this appendix, I describe my conventions and show some of the details 
of calculations described in the main body of the paper.

\subsection{Field equations}

As in the main text, I work in a Poincar{\'e} coordinate patch, with 
$\Lambda=-1/\ell^2 = -1$.   My metric signature is $-++$, with 
$\tilde\epsilon^{+-z} = \sqrt{|g|}\epsilon^{+-z} = -1$.
As in \cite{CDWW}, I use the ``wrong sign'' Newton's constant $G<0$,
which connects smoothly to the standard choice at $\Lambda=0$.
For solutions of the field equations, changing the sign of $G$ is 
equivalent to reversing chirality, so my ``$+$'' components are the 
``$-$'' components of  \cite{LSS,MSS}.  

In the Fefferman-Graham gauge 
(\ref{a6}), the linearization of the cosmological Einstein  tensor $\G_{\mu\nu} 
= G_{\mu\nu} - g_{\mu\nu}$ around anti-de Sitter space is
\begin{alignat}{2}
\HH_{++} &= -\frac{1}{2}N(N+2)h_{++} \qquad\qquad &&  
\HH_{+z} = \frac{z}{2}(N+2)(\partial_-h_{++} - \partial_+h_{+-}) 
    \nonumber\\
\HH_{+-} &= \frac{1}{2}N(N+2)h_{+-} &&
 \HH_{-z} = \frac{z}{2}(N+2)(\partial_+h_{--} - \partial_-h_{+-}) 
\label{Aa1} \\
\HH_{--} &= -\frac{1}{2}N(N+2)h_{--} &&
\HH_{zz} =-(N+2)h_{+-} -\frac{z^2}{2}\left( 
    \partial_-^2h_{++} - 2\partial_+\partial_-h_{+-} + \partial_+^2h_{--}
    \right) . \nonumber 
\end{alignat}
The linearized field equations (\ref{c2}) are
\begin{align}
\HH_{\mu\nu} 
  + \epsilon_\mu{}^{\alpha\beta}{\bar\nabla}_\alpha\HH_{\beta\nu} = 0 .
\label{Aa2}
\end{align}
(The second term may be symmetrized in $\mu$ and $\nu$, but this need 
not be done explicitly, since the antisymmetric part vanishes by virtue of 
the Bianchi identities.)  The six independent components of (\ref{Aa2}) 
may be taken to be
\begin{align} 
&E_{++}^{(1)} = z\partial_+\HH_{z+} - N\HH_{++}
   = \frac{z^2}{2}(N+2)(\partial_+\partial_-h_{++} - \partial_+{}^2h_{+-})
  + \frac{1}{2}N^2(N+2)h_{++} \nonumber\\
&E_{+-}^{(1)}- {\cal E}
   = -z\partial_-\HH_{z+} + N\HH_{+-} 
   = - \frac{z^2}{2}(N+2)(\partial_-{}^2h_{++} - \partial_+\partial_-h_{+-})
   + \frac{1}{2}N^2(N+2)h_{+-} \nonumber\\
&E_{--}^{(1)} = -z\partial_-\HH_{z-} + (N+2)\HH_{--} 
   = -\frac{z^2}{2}(N+2)(\partial_+\partial_-h_{--} - \partial_-{}^2h_{+-})
   - \frac{1}{2}N(N+2)^2h_{--} \nonumber\\
&E_{+z}^{(1)} = z(\partial_-\HH_{++} - \partial_+\HH_{+-})
   = -\frac{z}{2}N(N+2)(\partial_-h_{++} + \partial_+h_{+-})   
\label{Aa3} \\
&E_{-z}^{(1)} + z\partial_-{\cal E} 
   = 2z\partial_-\HH_{+-} + (N+1)\HH_{-z} 
   = \frac{z}{2}\left[ (N+2)^2\partial_+h_{--} + (N-2)(N+2)\partial_-h_{+-}
   \right] \nonumber\\
&{\cal E} = 2\HH_{+-} + \HH_{zz}
    = (N-1)(N+2)h_{+-} - \frac{z^2}{2}(\partial_-{}^2h_{++} 
    - 2\partial_+\partial_-h_{+-} + \partial_+{}^2h_{--}) . \nonumber
\end{align}
where $N=z\partial_z$ and ${\cal E} = {\bar g}^{\mu\nu}E_{\mu\nu}^{(1)}$.

\subsection{General solution}

Our next goal is to find the general solution of (\ref{Aa2}).  Without 
loss of generality, we may parametrize the metric perturbation 
$h_{+-}$ as
\begin{align}
h_{+-} = 2(N+2)\partial_+\partial_-Z  ,
\label{Ab1}
\end{align}
where $Z$ is an arbitrary function of $x^+$, $x^-$, and $z$.  Then
\begin{align} 
    E_{-z}^{(1)}=0 \quad&\Rightarrow 
    (N+2)^2\partial_+\left[h_{--} + 2 (N-2)\partial_-{}^2Z\right]=0 
    \nonumber\\
    &\Rightarrow h_{--} = -2(N-2)\partial_-{}^2Z + A \quad
    \hbox{with $(N+2)^2\partial_+A=0$} ,
\label{Ab2}\\
    E_{+z}^{(1)}=0 \quad&\Rightarrow 
    N(N+2)\partial_-\left[h_{++} + 2 (N+2)\partial_+{}^2Z\right]=0 
    \nonumber\\
    &\Rightarrow h_{++} = -2(N+2)\partial_+{}^2Z + B \quad
    \hbox{with $N(N+2)\partial_-B=0$} . \nonumber
\end{align}
The conditions on the functions $A$ and $B$ require that
\begin{align}
A &= \frac{\partial_-{}^2a_1}{z^2} 
      + \frac{\partial_-{}^2a_2\ln z}{z^2} + v^-(x^-,z) 
\label{Ab3}\\
B &= \frac{\partial_-{}^2b_1}{z^2} 
      + \partial_+{}^3\partial_-b_2 + v^+(x^+,z)  \nonumber
\end{align}
where $a_1$, $a_2$, $b_1$, and $b_2$ are arbitrary functions of $x^+$ 
and $x^-$; the derivatives $\partial_\pm$ are inserted for later 
notational convenience, and do not affect the generality of the solution.   
The equations ${\cal E}^{(1)}=0$ and $E_{+-}^{(1)}=0$ then become
\begin{align}
(N-1)\partial_+\partial_-Y = \frac{z^2}{4}(\partial_-{}^2B 
      + \partial_+{}^2A) ,\qquad
N^2\partial_+\partial_-Y = \frac{z^2}{2}(N+2)\partial_-{}^2B  
\label{Ab4}
\end{align}
with
\begin{align}
Y = 2z^2\partial_+\partial_-Z + (N+2)^2Z   .
\label{Ab5}
\end{align}
These are straightforward to integrate, yielding
\begin{align} 
Y = \frac{z^2}{4}\partial_+{}^2\partial_-{}^2b_2 
    - \frac{1}{4}\partial_+\partial_-(a_1+a_2+b_1)
    + (N+2)^2w^+(x^+,z) + (N+2)^2w^-(x^-,z)
\label{Ab6}
\end{align}
where the $w^\pm$ are arbitrary functions of their arguments.  The 
remaining field equations $E_{++}^{(1)} = E_{--}^{(1)} = 0$ then reduce to
\begin{align}
N^2(N+2)\left[ v^+ - 2(N+2)\partial_+{}^2w^+\right] = 0 &\Rightarrow
    v^+ - 2(N+2)\partial_+{}^2w^+ = \frac{\alpha^+}{z^2} + \beta^+ 
    + \gamma^+\ln z
 \nonumber\\
N(N+2)^2\left[ v^- - 2(N-2)\partial_-{}^2w^-\right] = 0  &\Rightarrow
    v^- - 2(N-2)\partial_-{}^2w^- = \frac{\alpha^-}{z^2} + \beta^- 
   + \frac{\gamma^-\ln z}{z^2}
\label{Ab7} 
\end{align}
where $(\alpha^+,\beta^+,\gamma^+)$ and $(\alpha^-,\beta^-,\gamma^-)$
are arbitrary functions of $x^+$ and $x^-$, respectively.

Our next step is to solve (\ref{Ab5}) for $Z$, given the source (\ref{Ab6}).  
It is easy to check that a particular solution is
\begin{align}
Z_0 = -\frac{a_1+a_2+b_1+2b_2}{8}\cdot\frac{1}{z^2} 
        - \frac{a_2}{8}\cdot\frac{\ln z}{z^2}
       + \frac{1}{8}\partial_+\partial_-b_2 + w^+ + w^- .
\label{Ab8}
\end{align}
The general solution will be of the form 
\begin{align}
Z=Z_0+X ,
\label{Ab9}
\end{align}
where $X$ is a solution of the homogeneous equation (\ref{a5}).

Finally, we insert (\ref{Ab3}), (\ref{Ab7}), (\ref{Ab8}), and (\ref{Ab9})  
into (\ref{Ab1})--(\ref{Ab3}) to determine the full first-order metric 
perturbation.   A straightforward computation now yields the result 
(\ref{a7})--(\ref{a8}), where $X$ and $\gamma^+$ in (\ref{a7}) and 
(\ref{a8})  are as in (\ref{Ab9}) and (\ref{Ab7}), and  $a$, $b$, and $c$ 
in (\ref{a8}) are linear combinations of $a_1$, $a_2$, $b_1$, and $b_2$ 
in (\ref{Ab3}).

\subsection{Linearization instability}

I next turn to the derivation of equation (\ref{c4}).  I will make liberal use of 
the identity, true for any $B_{\mu\nu}$, that
$$B_{\mu\nu} - B_{\nu\mu} 
  = -\epsilon_{\mu\nu\rho}\epsilon^{\rho\sigma\tau}B_{\sigma\tau} $$
(where the sign on the right-hand side follows from my ``mostly minus''
metric signature convention).  We then have
\begin{align}
\xi^\mu E_\mu{}^\nu = &-\chi^\mu\G_\mu{}^\nu 
   -\frac{1}{2}\epsilon^{\mu\alpha\beta}\nabla_\alpha\xi_\beta\G_\mu{}^\nu
   +\frac{1}{2}\epsilon^{\mu\alpha\beta}\xi_\mu\nabla_\alpha\G_\beta{}^\nu
   +\frac{1}{2}\epsilon^{\nu\alpha\beta}\xi_\mu\nabla_\alpha\G_\beta{}^\mu
\nonumber\\
   = &-\chi^\mu\G_\mu{}^\nu 
   -\frac{1}{2}\nabla_\alpha
         \left(\epsilon^{\mu\alpha\beta}\xi_\beta\G_\mu{}^\nu 
         - \epsilon^{\mu\nu\beta}\xi_\beta\G_\mu{}^\alpha\right)\nonumber\\
    &-\frac{1}{2}\epsilon^{\mu\nu\beta}\nabla_\alpha\xi_\beta\G_\mu{}^\alpha 
    +\frac{1}{2}\epsilon^{\mu\alpha\beta}
          \left(\xi_\beta\nabla_\alpha\G_\mu{}^\nu
         +\xi_\mu\nabla_\alpha\G_\beta{}^\nu\right)
    +\frac{1}{2}\epsilon^{\nu\alpha\beta}\xi_\mu\nabla_\alpha\G_\beta{}^\mu 
\label{Ac1}\\
    = &-\chi^\mu\G_\mu{}^\nu 
    -\frac{1}{2}\nabla_\alpha\left(
       \epsilon^{\mu\alpha\beta}\xi_\beta\G_\mu{}^\nu 
       - \epsilon^{\mu\nu\beta}\xi_\beta\G_\mu{}^\alpha 
       - \epsilon^{\nu\alpha\beta}\xi_\mu\G_\beta{}^\mu\right) \nonumber\\
    &-\frac{1}{2}\left(
        \epsilon^{\alpha\nu\beta}\nabla_\mu\xi_\beta\G_\alpha{}^\mu
       + \epsilon^{\nu\alpha\beta}\nabla_\alpha\xi_\mu\G_\beta{}^\mu\right) 
    = -\chi^\mu\G_\mu{}^\nu 
       + \frac{1}{2}\epsilon^{\nu\alpha\beta}\Delta_{\mu\beta}\G_\alpha{}^\mu
       +\nabla_\alpha{\cal F}^{\alpha\nu} \nonumber
\end{align}
with
\begin{align}
{\cal F}^{\alpha\nu} &= -\frac{1}{2}\left(
         \epsilon^{\mu\alpha\beta}\xi_\beta\G_\mu{}^\nu 
         - \epsilon^{\mu\nu\beta}\xi_\beta\G_\mu{}^\alpha 
         - \epsilon^{\nu\alpha\beta}\xi_\mu\G_\beta{}^\mu\right)  
   = \frac{1}{2}\epsilon^{\alpha\nu\sigma}\epsilon_{\sigma\lambda\tau}
         \epsilon^{\mu\lambda\beta}\xi_\beta\G_\mu{}^\tau 
        + \frac{1}{2}\epsilon^{\nu\alpha\beta}\xi_\mu\G_\beta{}^\mu  
        \nonumber \\
   &= - \frac{1}{2}\epsilon^{\alpha\nu\sigma}\delta^{\mu\beta}_{\sigma\tau}
        \xi_\beta\G_\mu{}^\tau
        + \frac{1}{2}\epsilon^{\nu\alpha\beta}\xi_\mu\G_\beta{}^\mu  
      = \epsilon^{\nu\alpha\beta}\xi_\nu\left(\G_\beta{}^\mu 
         - \frac{1}{2}\delta^\mu_\beta\G\right) .
   \label{Ac2}
\end{align}
I next turn to equation (\ref{c9}).   Note first that from (\ref{c7a})
\begin{align}
(\sqrt{|g|}\,\chi^+)^{(1)} &= -\frac{1}{2}\frac{1}{z}(N+2)h_{+-} \nonumber\\
(\sqrt{|g|}\,\chi^-)^{(1)} &= \frac{1}{2}\frac{1}{z}N\,h_{++} \label{Ac2a}\\
(\sqrt{|g|}\,\chi^-)^{(1)} &= \frac{1}{2}(\partial_+h_{+-} - \partial_-h_{++}) . 
\nonumber
\end{align}
Let us start by considering the quantity $\sqrt{|g|}\,\xi^\mu E_\mu^{(2)-}$.  
From (\ref{c7a}) and (\ref{Aa1}), the terms on the right-hand side of (\ref{c7}) are
\begin{align}
-z^2(\sqrt{|g|}\,\chi^\mu)^{(1)}\HH_{\mu +} 
  = &-\frac{z}{4}(N+2)h_{+-}\cdot(N+2)Nh_{++} 
         - \frac{z}{4}N(N+2)h_{+-}\cdot Nh_{++}\nonumber\\
      &+ \frac{z^3}{4}(\partial_+h_{+-}-\partial_-h_{++})
         (N+2) (\partial_+h_{+-}-\partial_-h_{++})
\label{Ac3}\\
  = & -\partial_z\left[ \frac{z^2}{4}(N+2)h_{+-}\cdot Nh_{++}
     - \frac{z^4}{8}\left(\partial_+h_{+-}-\partial_-h_{++}\right)^2\right] \nonumber \\
-\frac{z^2}{2}\Delta^{(1)}_{+-}\HH_{z+}
  = & -\frac{z^3}{4}\partial_+h_{+-}(N+2)(\partial_-h_{++} - \partial_+h_{+-})
\label{Ac4}\\
  = &\ \partial_z\left[ \frac{z^4}{4}\left(\partial_+h_{+-}\right)^2\right] 
     - \frac{z^3}{4}\partial_+h_{+-}(N+2)(\partial_-h_{++} + \partial_+h_{+-})
     \nonumber \\
-\frac{z^2}{2}\Delta^{(1)}_{++}\HH_{z-}
  = & -\frac{z^2}{2}\partial_+(h_{++}\HH_{z-}) 
      + \frac{z}{4}h_{++}(N-2)N(N+2)h_{+-}\nonumber\\
  = & -\frac{z^2}{2}\partial_+(h_{++}\HH_{z-}) 
  - \frac{z}{4}h_{+-}N(N+2)(N+4)h_{++} 
\label{Ac5}\\ 
     &+ \partial_z\!\left[\frac{z^2}{4}\left( h_{++}(N-2)Nh_{+-}
      - Nh_{++}(N-2)h_{+-} +h_{+-} N(N+2)h_{++}\right)\right] \nonumber
\end{align}
where I have used the equation of motion $\partial_+\HH_{z-} = 
\frac{1}{z}(N-2)\HH_{+-}$ in the first line.  Now let
\begin{align}
V =& -\frac{z^2}{2}h_{++}\HH_{z-} - \frac{z^3}{2}h_{+-}(N+6)\partial_+h_{+-} 
   \nonumber\\
\partial_+V =& -\frac{z^2}{2}\partial_+(h_{++}\HH_{z-}) 
   - \frac{z^3}{2}\partial_+h_{+-}(N+6)\partial_+h_{+-} 
\label{Ac6}\\
   &- \frac{z^3}{2}(N+6)\left(\partial_+{}^2h_{+-} 
              - \frac{1}{2z^2}N(N+2)h_{++}\right)h_{+-} 
      - \frac{z}{4}h_{+-}N(N+2)(N+4)h_{++} \nonumber\\
   =& -\frac{z^2}{2}\partial_+(h_{++}\HH_{z-}) 
      -\partial_z\left[\frac{z^4}{4}\left(\partial_+h_{+-}\right)^2\right]
      - z^3\left(\partial_+h_{+-}\right)^2 \nonumber\\
  &- \frac{z^3}{2}h_{+-} (N+6)\left(\partial_+{}^2h_{+-} 
              - \frac{1}{2z^2}N(N+2)h_{++}\right) 
      - \frac{z}{4}h_{+-}N(N+2)(N+4)h_{++} . \nonumber
\end{align}
Inserting (\ref{Ac3})--(\ref{Ac6}) into (\ref{c7}), we see that
\begin{align}
\sqrt{|g|}\,&\xi^\mu E_\mu^{(2,\text{bulk})-} = z^3\left(\partial_+h_{+-}\right)^2
    +\partial_+V + \partial_zU^- \label{Ac7}\\
    &+ \frac{z^3}{2}h_{+-} (N+6)\left(\partial_+{}^2h_{+-} 
              - \frac{1}{2z^2}N(N+2)h_{++}\right) 
     - \frac{z^3}{4}\partial_+h_{+-}(N+2)(\partial_-h_{++} + \partial_+h_{+-}) ,
     \nonumber
\end{align}
where $V$ is as in (\ref{Ac6}) and  
\begin{align}
U^- =&\ \frac{z^4}{2}(\partial_+h_{+-})^2 
             + \frac{z^4}{8}(\partial_+h_{+-}-\partial_-h_{++})^2 \nonumber\\
             &+ \frac{z^2}{4}\left( h_{++}(N-2)Nh_{+-} -2 Nh_{++}Nh_{+-}
             + h_{+-}N(N+2)h_{++}\right) .
\label{Au1}
\end{align}
We next turn to $\sqrt{|g|}\,\xi^\mu E_\mu^{(2)+}$, which by (\ref{c7}) 
and (\ref{c7a}) is
\begin{align}
\sqrt{|g|}\,\xi^\mu E_\mu^{(2,\text{bulk})+} = \frac{z}{2}(N+2)h_{+-}\HH_{+-}
  - \frac{z}{2}Nh_{++}\HH_{--} + \frac{z^2}{2}\partial_-h_{++}\HH_{z-}
  + \frac{z^2}{2}\partial_+h_{--}\HH_{z+} .
\label{Ac8}
\end{align}
It is then straightforward to see from (\ref{Ac6}) that
\begin{align} 
\sqrt{|g|}\,&\xi^\mu E_\mu^{(2,\text{bulk})+} = - \partial_-V
  + \frac{z}{4}(N+2)h_{+-}N(N+2)h_{+-} 
  - \frac{z^3}{2}h_{+-}(N+6)\partial_+\partial_-h_{+-} \nonumber\\
  &\ - \frac{z^3}{2}\partial_-h_{+-}(N+6)\partial_+h_{+-}
  + \frac{z^3}{4}\partial_+h_{--}(N+2)(\partial_-h_{++} - \partial_+h_{+-})
  - \partial_z\left[\frac{z^2}{2}h_{++}\HH_{--}\right] \nonumber\\
  =&\ - \partial_-V + zh_{+-}N(N+2)h_{+-} 
     + \frac{z^3}{2}(N-2)h_{+-}\left[\partial_+\partial_-h_{+-}
     + \frac{1}{2z^2}N(N+2)h_{+-}\right] \nonumber\\
  &+ \frac{z^3}{2}\left[ (N-2)\partial_-h_{+-} 
             + (N+2)\partial_+h_{--}\right]\partial_+h_{+-}  
   +\frac{z^3}{4}\partial_+h_{--}(N+2)(\partial_-h_{++} + \partial_+h_{+-})
   \nonumber\\
   &- \partial_z\!\left[\frac{z^2}{2}h_{++}\HH_{--} 
             + \frac{z^4}{4}\left(h_{+-}\partial_+\partial_-h_{+-} 
             + \partial_+h_{+-}\partial_-h_{+-} 
             + \partial_+h_{--}\partial_+h_{+-}\right)\right] 
\label{Ac9} \\
  =&\ - \partial_-V - \partial_zU^+ - z\left[(N+2)h_{+-} \right]^2
     + \frac{z^3}{2}(N-2)h_{+-}\left[\partial_+\partial_-h_{+-}
     + \frac{1}{2z^2}N(N+2)h_{+-}\right] \nonumber\\
  &+ \frac{z^3}{2}\left[ (N-2)\partial_-h_{+-} 
             + (N+2)\partial_+h_{--}\right]\partial_+h_{+-}  
   +\frac{z^3}{4}\partial_+h_{--}(N+2)(\partial_-h_{++} + \partial_+h_{+-})
   \nonumber
\end{align}
with
\begin{align}
U^+ = \frac{z^2}{2}h_{++}\HH_{--} - z^2h_{+-}(N+2)h_{+-}
             + \frac{z^4}{4} \left(h_{+-}\partial_+\partial_-h_{+-} 
             +  \partial_+h_{+-}\partial_-h_{+-} 
             +  \partial_+h_{--}\partial_+h_{+-}\right) .
\label{Au2}
\end{align}

Now recall first that at first order, $h$ has two contributions (\ref{a7}) 
and (\ref{a8}).  The integral $I_\xi^{(2)}$ will correspondingly have 
three contributions: one quadratic in $X$, one quadratic in $\tilde h$, 
and one cross term.  Let us first consider the term quadratic in $X$.
From (\ref{a7}), 
\begin{align}
\partial_+{}^2h_{+-}  - \frac{1}{2z^2}N(N+2)h_{++} &= 0  \nonumber\\
\partial_-h_{++} + \partial_+h_{+-} &= 0 \label{Ac10}\\
\partial_+\partial_-h_{+-} + \frac{1}{2z^2}N(N+2)h_{+-} &= 0 \nonumber\\
(N-2)\partial_-h_{+-} + (N+2)\partial_+h_{--} &= 0 . \nonumber
\end{align}
Inserting (\ref{Ac10}) into (\ref{Ac7}) and (\ref{Ac9}) and combining to
form the $t$ component $\sqrt{|g|}\,\xi^\mu E_\mu^{(2,\text{bulk})t}$, 
 we immediately obtain (\ref{c9}).

We next consider the contributions involving $\tilde h$.  We could again
use (\ref{Ac7}) and (\ref{Ac9}), but it is simpler to return to (\ref{c7}).
Observe that the only nonvanishing contribution of $\tilde h$ to the
linearized curvature is $\widetilde\HH_{++} = -\gamma^+$, so (\ref{c7}) 
will consist almost entirely of cross terms involving $\HH[X]$.  Note also
that if $f(x^+,x^-)$ is any $z$-independent function, then it follows from
(\ref{Aa1}) that $zf\HH_{++}$, $zf\HH_{+-}$, $zf\HH_{--}$, $z^{-1}f\HH_{+-}$, 
$z^{-1}f\HH_{--}$, $f\HH_{z+}$, and $f\HH_{z-}$ are all total $z$ derivatives.  
It is then easy to check that the only remaining $\tilde h$-dependent
contributions to $\sqrt{|g|}\,\xi^\mu E_\mu^{(2)+}$ are
\begin{align}
\sqrt{|g|}\,\xi^\mu E_\mu^{(2,\text{bulk})+} = \dots
   + \frac{z^2}{8}\left[ \partial_+{}^3\partial_-{}^2(2a-b-c)\HH_{z-}
   + \partial_+{}^2\partial_-{}^3(2a-b-c)\HH_{z+}\right] .
\label{Ac11}
\end{align}
But
\begin{align}
\HH_{z-} &= \frac{z}{2}(N+2)(\partial_+h_{--} - \partial_-h_{+-})
    = -2zN(N+2)\partial_+\partial_-{}^2X = \frac{1}{z}(N-2)N(N+2)^2\partial_-X
\nonumber\\
\HH_{z+} &= \frac{z}{2}(N+2)(\partial_-h_{++} - \partial_+h_{+-})
    = -2z(N+2)^2\partial_+{}^2\partial_-X = \frac{1}{z}N^2(N+2)^2\partial_+X ,
\label{Ac12}
\end{align}
so the terms in (\ref{Ac11}) are each of the form 
$$zf(x^+,x^-)(N+2)g = \partial_z(z^2fg).$$ 

Similarly, the only $\tilde h$-dependent contributions to 
$\sqrt{|g|}\,\xi^\mu E_\mu^{(2)-}$ that are not immediately recognizable
as total $z$ derivatives are
\begin{align}
\sqrt{|g|}\,\xi^\mu E_\mu^{(2,\text{bulk})-} = \dots
   &- \frac{z^2}{8}\left[ \partial_+{}^4\partial_-(2a-b-c)\HH_{z-}
   + \partial_+{}^3\partial_-{}^2(2a-b-c)\HH_{z+}\right] \nonumber\\
   &- \frac{z^2\ln z}{2}\partial_+\gamma^+\HH_{z-} 
   - \frac{z}{2}\gamma^+(N+2)h_{+-}  . \label{Ac13}
\end{align}
As in (\ref{Ac11}), the first two terms are again total $z$ derivatives, as
is the last term.  The remaining term is also a total derivative, although 
less obviously:
\begin{align}
- \frac{z^2\ln z}{2}&\partial_+\gamma^+\HH_{z-} = 
   - \frac{z\ln z}{2}\partial_+\gamma^+ (N-2)N(N+2)^2\partial_-X \label{Ac14}\\
   &= -\partial_z\left[ \frac{z^2\ln z}{2}\partial_+\gamma^+
         (N-2)N(N+2)\partial_-X\right] 
         + \frac{z}{2}\partial_+\gamma^+(N-2)N(N+2)\partial_-X \nonumber\\
    &= -\partial_z\left[ \frac{z^2\ln z}{2}\partial_+\gamma^+
         (N-2)N(N+2)\partial_-X - \frac{z^2}{2}\partial_+\gamma^+
         (N-2)N\partial_-X\right] . \nonumber
\end{align}
We thus conclude that as long as $X$ has compact support, the terms
involving $\tilde h$ make no contribution to the integral $I_\xi^{(2)}$, and
the expression (\ref{c9}) is fully general.

\vspace{1.5ex}
\begin{flushleft}
\large\bf Acknowledgments
\end{flushleft}

I would like to thank my colleagues Geoffrey Compere, Stanley Deser, 
Alex Maloney, Don Marolf, Kristen Schleich, Wei Song, Andy Strominger, 
Andrew Waldron, Derek Wise, and Don Witt for helpful discussions about 
issues treated here.  I am grateful for the hospitality of the Erwin 
Schr{\"o}dinger Institute, where a portion of this research was 
completed.  This work was supported in part by U.S.\ Department of Energy 
grant DE-FG02-91ER40674.

\end{document}